\documentclass[notitlepage,aps,prb,twocolumn,groupedaddress,showpacs,superscriptaddress,10pt]{revtex4-1}

\usepackage[Symbolsmallscale]{upgreek}
\usepackage{lmodern}
\usepackage[T1]{fontenc}

\usepackage{amsmath,amssymb,mathrsfs,graphicx,comment}
\usepackage[dvipsnames,usenames]{color}
\usepackage{bm}
\usepackage{ulem}
\usepackage{textcomp}
\usepackage{pbox}

\begin{document}
\title{Machine learning applied to single-shot x-ray diagnostics in an XFEL}

\newcommand{\ImperialPhysics}{Department of Physics, Imperial College, London, SW7 2AZ, United Kingdom}
\newcommand{\LCLS}{Linac Coherent Light Source, SLAC National Accelerator Laboratory, Menlo Park, California 94025, USA}
\newcommand{\Gothenburg}{Department of Physics, University of Gothenburg, Origov{\"a}gen 6B, 41296 Gothenburg, Sweden}
\newcommand{\UConn}{Department of Physics, University of Connecticut, 2152 Hillside Road, U-3046, Storrs, CT 06269, USA}
\newcommand{\Lutheran}{California Lutheran University, 60 W Olsen Rd, Thousand Oaks, CA 91360, USA}
\newcommand{\Stanford}{Department of Physics, Stanford University, 382 Via Pueblo Mall, Stanford, CA 94305, USA}
\newcommand{\ImperialChem}{Department of Chemistry, Imperial College London, London SW7 2AZ, United Kingdom}
\newcommand{\UppsalaChemistry}{Department of Chemistry - {\AA}ngtr{\"o}m, Uppsala University, Uppsala, 75120, Sweden}
\newcommand{\Tohoku}{Institute of Multidisciplinary Research for Advanced Materials, Tohoku University, Sendai 980-8577, Japan}
\newcommand{\Argonne}{Argonne National Laboratory, Lemont, IL 60439, USA}
\newcommand{\MAXVI}{MAX IV Laboratory, Lund University, Box 118, SE-221 00 Lund, Sweden}
\newcommand{\UppsalaPhysics}{Department of Physics and Astronomy, Uppsala University, Uppsala, 75120,  Sweden}
\newcommand{\XFEL}{European XFEL GmbH, Holzkoppel 4, 22869 Schenefeld, Germany}
\newcommand{\PULSE}{Stanford PULSE Institute, SLAC National Accelerator Laboratory, Menlo Park, California 94025, USA}
\newcommand{\DESY}{Deutsches Elektronen-Synchrotron DESY, Notkestrasse 85, Hamburg, 22607, Germany}
\newcommand{\Kassel}{Institut f{\"u}r Physik und CINSaT, Universit{\"a}t Kassel, Heinrich-Plett-Str. 40, 34132 Kassel, Germany}
\newcommand{\TUM}{Physics Department, TU Munich, James-Franck-Str. 1, 85748 Garching, Germany}
\newcommand{\Qamcom}{Qamcom Research & Technology AB, Falkenbergsgatan 3, SE-412 85 G{\"o}teborg, Sweden}

\author{A. Sanchez-Gonzalez}\affiliation{\ImperialPhysics}
\author{P. Micaelli}\affiliation{\ImperialPhysics}
\author{C. Olivier}\affiliation{\ImperialPhysics}
\author{T. R. Barillot}\affiliation{\ImperialPhysics}
\author{M. Ilchen}\affiliation{\PULSE}\affiliation{\XFEL}
\author{A. A. Lutman}\affiliation{\LCLS}
\author{A. Marinelli}\affiliation{\LCLS}
\author{T. Maxwell}\affiliation{\LCLS}
\author{A. Achner}\affiliation{\XFEL}
\author{M. Ag{\aa}ker}\affiliation{\UppsalaPhysics}
\author{N. Berrah}\affiliation{\UConn}
\author{C. Bostedt}\affiliation{\LCLS}\affiliation{\Argonne}
\author{J. Buck}\affiliation{\DESY}
\author{P. H. Bucksbaum}\affiliation{\PULSE}\affiliation{\Stanford}
\author{S. Carron Montero}\affiliation{\LCLS}\affiliation{\Lutheran}
\author{B. Cooper}\affiliation{\ImperialPhysics}
\author{J. P. Cryan}\affiliation{\PULSE}
\author{M. Dong}\affiliation{\UppsalaPhysics}
\author{R. Feifel}\affiliation{\Gothenburg}
\author{L. J. Frasinski}\affiliation{\ImperialPhysics}
\author{H. Fukuzawa}\affiliation{\Tohoku}
\author{A. Galler}\affiliation{\XFEL}
\author{G. Hartmann}\affiliation{\DESY}\affiliation{\Kassel}
\author{N. Hartmann}\affiliation{\LCLS}
\author{W. Helml}\affiliation{\LCLS}\affiliation{\TUM}
\author{A. S. Johnson}\affiliation{\ImperialPhysics}
\author{A. Knie}\affiliation{\Kassel}
\author{A. O. Lindahl}\affiliation{\PULSE}\affiliation{\Gothenburg}
\author{J. Liu}\affiliation{\XFEL}
\author{K. Motomura}\affiliation{\Tohoku}
\author{M. Mucke}\affiliation{\UppsalaPhysics}
\author{C. O'Grady}\affiliation{\LCLS}
\author{J-E. Rubensson}\affiliation{\UppsalaPhysics}
\author{E. R. Simpson}\affiliation{\ImperialPhysics}
\author{R. J. Squibb}\affiliation{\Gothenburg}
\author{C. S{\aa}the}\affiliation{\MAXVI}
\author{K. Ueda}\affiliation{\Tohoku}
\author{M. Vacher}\affiliation{\ImperialChem}\affiliation{\UppsalaChemistry}
\author{D. J. Walke}\affiliation{\ImperialPhysics}
\author{V. Zhaunerchyk}\affiliation{\Gothenburg}
\author{R. N. Coffee}\affiliation{\LCLS}
\author{J. P. Marangos}\affiliation{\ImperialPhysics}

\begin{abstract}
X-ray free-electron lasers (XFELs) are the only sources currently able to produce bright few-fs pulses with tunable photon energies from 100 eV to more than 10 keV. Due to the stochastic SASE operating principles and other technical issues the output pulses are subject to large fluctuations, making it necessary to characterize the x-ray pulses on every shot for data sorting purposes. We present a technique that applies machine learning tools to predict x-ray pulse properties using simple electron beam and x-ray parameters as input. Using this technique at the Linac Coherent Light Source (LCLS), we report mean errors below 0.3 eV for the prediction of the photon energy at 530 eV and below 1.6 fs for the prediction of the delay between two x-ray pulses. We also demonstrate spectral shape prediction with a mean agreement of 97\%. This approach could potentially be used at the next generation of high-repetition-rate XFELs to provide accurate knowledge of complex x-ray pulses at the full repetition rate.
\end{abstract}
\maketitle

\section{Introduction}\label{Section1}
X-ray free-electron lasers (XFELs)\cite{emma2010first,ishikawa2012compact,allaria2013two} are emerging as one of the most versatile tools in x-ray research, becoming widely used by the scientific community, as well as industry, in many fields including physics, chemistry, biology, and material science. Their brightness, coherence, tunability, and ability to generate pairs of few-fs multicolor pulses for pump-probe experiments\cite{lutman2013experimental,marinelli2015high,hara2013two,lutman2016fresh} make them ideal sources to perform diffract-before-destroy imaging\cite{chapman2011femtosecond}, resonant x-ray spectroscopy\cite{gawelda2015x}, and a range of time resolved measurements of picosecond to few-femtosecond dynamics in molecules and atoms\cite{glownia2010time,erk2014imaging,marangos2011introduction,liekhus2015ultrafast,ullrich2012free,ferguson2016transient,pic2016hetero}.

A drawback to XFELs is their current poor stability. XFELs are driven by single-pass electron linear accelerators (LINAC) typically several hundred meters in length. High density electron bunches are formed in an electron photoinjector, accelerated in radiofrequency (RF) cavities, compressed in magnetic chicanes, and finally driven through one or multiple undulator sections where the electrons emit coherent x-ray pulses due to Self Amplified Spontaneous Emission (SASE)\cite{bonifacio1984collective} (see Figure \ref{FigSetup}) or amplify an external laser seed in High Gain Harmonic Generation (HGHG) schemes\cite{yu1991generation}. Small fluctuations in, for example, the photoinjector drive laser, RF amplitudes or phases along the LINAC translate into fluctuations in the XFEL pulse properties. Furthermore, all existing XFEL machines operating at wavelengths shorter than 4 nm have additional fluctuations due to the stochastic character of the SASE startup process, and shows only partial longitudinal coherence across the XFEL pulse. As a result, even a perfectly stable electron beam will exhibit shot-to-shot fluctuations in the fine temporal structure of the x-ray pulse. For example, when using single-pulse SASE emission at the LINAC Coherent Light Source (LCLS), fluctuations driven primarily by the LINAC RF systems lead to photon energy jitter of 0.1\% to 0.5\% full width half maximum (FWHM), depending on the central wavelength. The energy jitter drives bunch compression jitter leading to pulse length fluctuations of \textasciitilde5\% and intensity fluctuations from 1\% to 10\%. These numbers are exacerbated in more advanced lasing schemes such as the twin bunch technique\cite{marinelli2015high} where two electron bunches are accelerated simultaneously to produce two pulses with variable time delay and photon energy separation. In this case the inter pulse delay jitter is in the order of 10 to 15 fs FWHM and the intensity can fluctuate much more widely (\textasciitilde20-100\%).

Often the only way around such instabilities is performing a full x-ray characterization for each XFEL shot, using a variety of detection methods such as gas detectors (total pulse energy), single-shot x-ray spectrometers (wavelength, spectral shape and even polarization), and transverse deflecting cavities for the spent electron bunches (pulse shape, pulse duration and delay between pulses). Based on these measurements, one can circumvent issues with instability by retaining only the events presenting certain pulse characteristics, or even exploiting the jitter to act as an effective power\cite{kimberg2016stimulated}, wavelength\cite{sanchez2015auger,kimberg2016stimulated} or delay scan\cite{harmand2013achieving} by sorting and binning the events according to those characteristics. Unfortunately, some diagnostics that intercept the full beam such as optical spectrometers are incompatible with some experimental setups, requiring the x-rays to be either sent to the diagnostic line or to the sample. Furthermore, many of those essential diagnostics will not be compatible with the next generation of XFELs driven by superconducting LINACs operating at MHz rates such as European XFEL\cite{altarelli2006european} or the LCLS-II\cite{lcls2015linac}. Simple shot-to-shot diagnostics such as electron bunch monitors (beam position, beam energy, peak current), x-ray gas detectors, or particle time-of-flight (TOF) detectors can in principle work at the increased repetition rate but this will not always be the case. In general, devices that require intercepting the full electron or x-ray beam with a solid, such as screens to measure the electron bunch phase space streaked by a transverse cavity or the time tool\cite{harmand2013achieving} for optical pulse synchronization, will not be able to cope with the thermal load at the high repetition rate. Moreover, devices that rely on a CCD array will not have in most cases the required bandwidth for readout and storage of the data at MHz rates. This means that any experiment requiring single-shot characterization will likely be limited to a much lower (\textless  1 kHz) repetition rate.

In this paper we propose a general method applicable at any XFEL facility to obtain typical complex shot-to-shot diagnostics or any other variable at a higher repetition rate than that allowed by the corresponding acquisition devices. Using data from the LCLS operating at 120 Hz as an example, we found that much of the information usually extracted from slow, complex diagnostics such as the pump-probe delay in the twin bunch mode, the photon energy, or even the spectral shape of the x-ray pulses, is strongly correlated to electron bunch and x-ray properties measured by fast diagnostics. While this correlations are driven by physical processes, performing accurate direct modelling of every experimental aspect in machines as complex as XFELs is not currently possible. As an alternative we use generic linear, quadratic and more complex but well known machine learning models such as Neural Networks (NN)\cite{cheng1994neural} or Support Vector Regression (SVR)\cite{smola2004tutorial} to describe the non-trivial hidden correlations between the fluctuations in the simple diagnostics and the fluctuations in the complex diagnostics. Similar approaches have been successfully used for feedback-loops at particle accelerator facilities\cite{edelen2016neural}. Furthermore, by including fast gas detectors, measuring the total x-ray energy, which is sensitive to SASE fluctuations, we can make predictions accounting for some of the stochastic jitter. These models can be fitted or \emph{trained} to predict the output of complex diagnostics, such as pump-probe delay, using a small amount of training data obtained for a fraction of the shots containing full diagnostics. After applying standard testing techniques and estimating the accuracy of the predictions, the models can be used to calculate the missing variables that could not be measured using complex diagnostics for all the remaining shots. This has the potential of lessening the load on the data stream requirements in existing machines, as well as providing full repetition rate information for even the most complex of the diagnostics at the new generation of XFELs. 

\section{Methods}\label{Section2}

\subsection{Experimental setup}\label{Section2a}
\begin{figure*}[t]
	\centering
\includegraphics[trim=0cm 0.10cm 0cm 0.2cm,clip,width=1\linewidth]{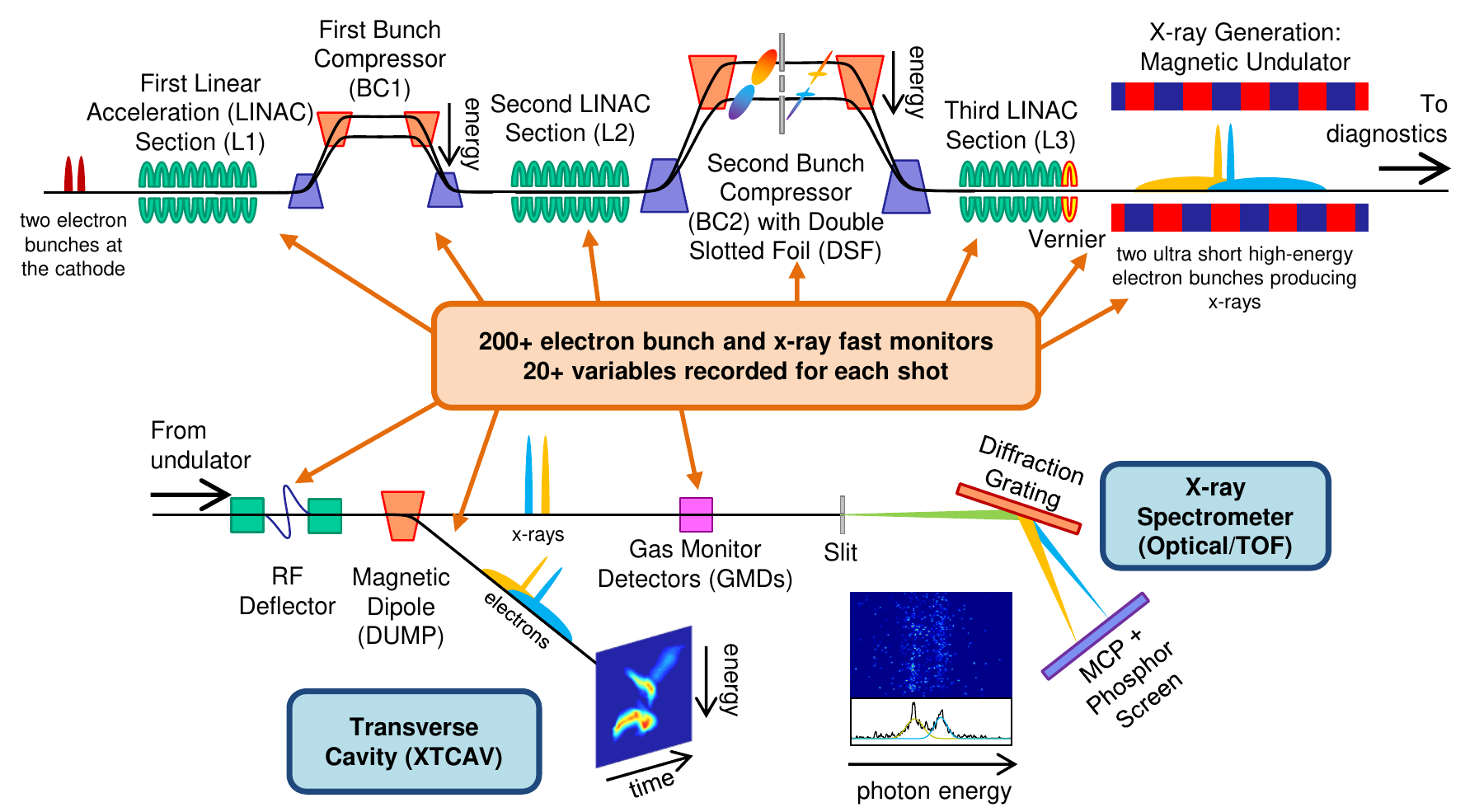}
	 \caption{\protect Schematic of the experimental setup. Many fast electron beam and x-ray detectors (orange box) are spread through all the XFEL sections. Critical single-shot data from slower detectors (blue boxes) is recorded at a lower repetition rate to train the machine learning models.}
	\label{FigSetup}
\end{figure*}

Experiments were conducted at the LCLS\cite{emma2010first} x-ray free-electron laser operated in the twin bunch mode\cite{marinelli2016twin} at the Atomic, Molecular and Optical Science (AMO)\cite{ferguson2015atomic} endstation in February (Expt. 1) and April (Expt. 2) of 2015. The experimental setup is depicted in Figure \ref{FigSetup}.

Two electron bunches were generated at 120 Hz at the photo-injector and accelerated in three different accelerator sections (L1,L2,L3), interleaved with two magnetic chicanes used as bunch compressors (BC), to energies near 3500 MeV and separated by 50 MeV\cite{emma2010first,marinelli2015high}. The resulting x-ray photon energies generated at the undulator were near the oxygen edge (540 eV) and separated by 15 eV. A double slotted foil (DSF) was used in the second chicane (BC2) to partially spoil each of the two electron bunches in time, limiting the emittance to a few femtoseconds duration\cite{emma2004femtosecond,ding2015generating}. By modifying bunch compression settings and the position of the DSF, it was possible to change the delay while maintaining the central photon energy of each of the pulses. We choose to demonstrate our technique in this mode of operation due to its versatility for ultrafast experiments, allowing two color, few-fs pulses, with an adjustable delay that can take any value from -100 fs and 100 fs, including zero-delay. The typical energies obtained for each pulse were spread over the range 0 to 30 \textmu J for the double pulse mode, and 15 to 45 \textmu J in the single pulse mode. All the data presented in this paper was taken at a fixed position of the foil and compression settings, with the different values for the time delay arising from fluctuations in the machine. In order to provide a larger range of photon energies, the final electron bunch energy was continuously scanned using the vernier.

An optical x-ray spectrometer (Expt. 1) and an electron TOF x-ray spectrometer\cite{allaria2014control} (Expt. 2) were used to measure single-shot spectra, each operating at 120 Hz. The optical x-ray spectrometer was calibrated using the absorption of a mylar filter\cite{rightor1997spectromicroscopy} at the oxygen K-edge and at the corresponding $\uppi^*$ resonance. The TOF x-ray spectrometer was calibrated using CO Auger electron emission at the oxygen K-edge and neon 2s and 2p photoelectrons at different photon energies. Under the applied experimental conditions, we found the signal-to-noise ratio of the optical spectrometer to be up to 16 times better than that of the TOF spectrometer, however, the average x-ray power in the double pulse mode during Expt. 2 was about three times larger, providing more signal.

An x-band transverse deflecting-mode cavity (XTCAV) was used to measure the single-shot spectrogram image of the electron bunches (time-energy distribution) downstream from the undulator at 60 Hz. By comparing images in the lasing and non-lasing cases one can determine the lasing region for each of the bunches and measure the distance along the time axis to obtain the pump-probe delay values\cite{ding2011femtosecond,behrens2014few,maxwell2014femtosecond} (Fig. \ref{FigDelay1}a-b).

Four gas detectors based on N$_2$ fluorescence\cite{hau2008measurement} were used to measure the single-shot total x-ray energy, recording 6 variables in total. Hundreds of different electron beam parameters were measured on each shot, however, only 16 of them were recorded at the full repetition rate. These included position monitors\cite{smith2009commissioning} (position and angle), bunch charge monitors, and peak current monitors at different stages (accelerators, chicanes, undulators) as is indicated in Figure \ref{FigSetup}. All these diagnostics consist of fast, non-intrusive detectors, and should be therefore scalable to the MHz regime.

Additionally nearly 300 ``slow'' variables were recorded at 2 Hz by the Experimental Physics and Industrial Control System (EPICS)\cite{dalesio1994experimental}. These variables mainly include temperatures of different sections or devices, pressures in the chambers, configuration values such as voltages or field strengths, and the settings of the many slow feedback loops that keep the FEL stable. The purpose of these variables was to monitor long term drifts, which can be useful to understand how the fluctuations evolve over time. More details about the variables included in the analysis can be found in Appendix 1.

\subsection{Computational methods}\label{Section2b}
\begin{figure*}[t]
	\centering
\includegraphics[trim=0cm 0cm 0cm 0cm,clip,width=0.75\linewidth]{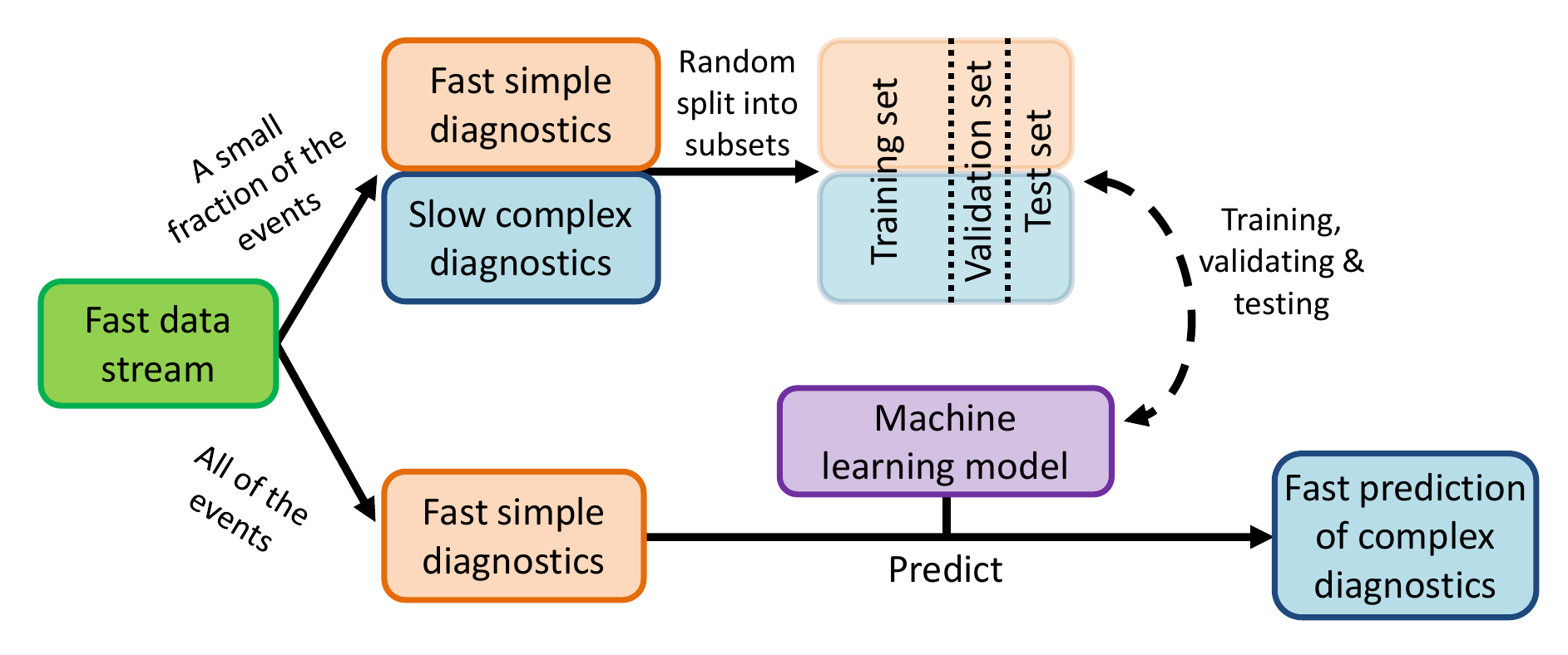}
	 \caption{\protect Schematic of the technique based on machine learning to predict complex diagnostics at a high repetition rate based on small samples of complex diagnostics obtained at a much lower repetition rate.}
	\label{FigML}
\end{figure*}

The proposed technique for the prediction of x-ray pulse characteristics at higher repetition rate than measured is summarized in Figure \ref{FigML}. It relies on a fast, high-repetition-rate data stream containing single-shot information of simple diagnostics for all the events, with information from complex diagnostics obtained at a lower repetition rate and only for a fraction of the events. The set of events containing correlated information from all devices can be split in three: the training, validation, and test sets. The training set is used to train a machine learning model to learn how to predict variables normally obtained with complex diagnostics, based on input variables from simple diagnostics. The validation set is used to optimize the \emph{hyperparameters}. In this context, a hyperparameter is any parameter of the model that is no optimized by the training process. Examples of hyperparameters are the maximum degree of a polynomial model, or the number of hidden layers in a NN. This optimization is done by training many different versions of the same model using different sets of hyperparameters, and then comparing the error on the validation set to decide which set of hyperparameters works best. Finally, the test set is used to test the prediction accuracy of the model for the chosen set of hyperparameters. At this point, the model can be applied to predict, with a known accuracy, the expected values from complex diagnostics for all the remaining events, which originally did not have that information.

We tested this approach using data from LCLS acquired at 60 Hz. It was implemented in python using the LCLS software package \emph{psana}\cite{damiani2016linac} at the LCLS servers, and locally on standard consumer computers. The Scikit-learn\cite{pedregosa2011scikit} framework was used for feature scaling, feature selection, Principal Component Analysis (PCA)\cite{jolliffe2002principal} and fitting of linear, polynomial and SVR (Gaussian kernel) models. Tensorflow\cite{abadi2016tensorflow} was used for the implementation of NNs.

More than 300 variables including signals from gas detectors, electron beam diagnostics, EPICS variables and a timestamp, were used as input variables or, in the common terminology of machine learning, \emph{features} for the prediction. More details about some of the particular variables included can be found in Appendix 1. The time delay, obtained from XTCAV, and the photon energy and spectral shape (\textasciitilde350 spectral components), measured with the spectrometers, were used as the output variables of the models. More details about each of these output variables can be found in the corresponding subsections for each of the prediction examples. As part of the feature selection process constant features were eliminated, as well as features taking a small number (\textless 10) of sparse discrete values. This normally reduced the total number of features to around 90. We then gradually reduced the number of features included, keeping only the ones showing a high correlation with the variable to be predicted, setting the threshold by minimizing the error of the validation set. Around 40 features were normally kept as a result of this process.

A typical dataset consisted of about $3\cdot 10^4$ shots. A filter was applied to remove all the shots where the total energy was below 5 \textmu J (\textless 10\% of all shots depending on the dataset). Shots presenting outliers in the outputs were also removed to avoid training on events where the results obtained from the complex diagnostics were potentially unreliable. We considered as outliers all the values separated from the median of the distribution by more than four times the median absolute deviation. This filtering process removed only a small fraction of the data (\textless 1\%), except in the spectral prediction case, where the noise due to single photon spikes in the optical spectrometer raised this value to 20\%. Each dataset was then divided randomly into 3 subsets using a common split for our dataset size, with 70\% of the data used for training, 15\% for validation and 15\% for testing. The test set was kept isolated from the rest during the training and optimization of the models.

\begin{table*}[ht]
\caption{\protect Summary of the mean error or agreement of the different prediction examples tested using the different models. The first column shows the mean error of the initial distribution. In the case of the shape agreement, this value corresponds to the mean agreement between each of the single-shot spectra and the mean spectrum. The values for each of the models correspond to the predictions on the test set, while the numbers in brackets correspond to the training set.}
\centering\begin{tabular}{c| c  c  c  c  c }
\pbox{20cm}{Test set (Train set)} & \pbox{20cm}{Initial\\Distribution} & \pbox{20cm}{Linear\\Model} & \pbox{20cm}{Quadratic\\Model} & \pbox{20cm}{Support Vector\\Regressor} & \pbox{20cm}{Neural\\Network} \\
\hline
\pbox{20cm}{Mean error of single\\pulse photon energy [eV]} & \pbox{20cm}{5.62} & \pbox{20cm}{0.29 (0.28)} & \pbox{20cm}{0.30 (0.24)} & \pbox{20cm}{0.32 (0.27)} & \pbox{20cm}{0.30 (0.29)} \\
\pbox{20cm}{Shape agreement of\\single pulse spectrum} & \pbox{20cm}{67\%} & \pbox{20cm}{88\% (88\%)} & \pbox{20cm}{94\% (95\%)} & \pbox{20cm}{95\% (95\%)} & \pbox{20cm}{97\% (97\%)} \\
\pbox{20cm}{Mean error of double\\pulse delay [fs]} & \pbox{20cm}{6.82} & \pbox{20cm}{2.07 (2.04)} & \pbox{20cm}{1.67 (1.58)} & \pbox{20cm}{1.67 (1.57)} & \pbox{20cm}{1.59 (1.52)} \\
\pbox{20cm}{Mean error of double\\pulse photon energy [eV]} & \pbox{20cm}{pulse 1: 1.45\\pulse 2: 1.03} & \pbox{20cm}{0.47 (0.49)\\0.44 (0.44)} & \pbox{20cm}{0.49 (0.48)\\0.41 (0.39)} & \pbox{20cm}{0.50 (0.48)\\0.41 (0.39)} & \pbox{20cm}{0.46 (0.47)\\0.40 (0.40)} \\
\end{tabular}
\label{TableSummary}
\end{table*}

Each of the features was normalized by subtracting the mean value and dividing on the standard deviation. This was also applied in some cases to the outputs, although we found the latter to only be relevant for the NNs. We did not find it necessary to use PCA on the features, as the total number of features feeded into the models was low (\textasciitilde40) for machine learning standards. On the other hand, we applied PCA to the output variables of the spectral shape prediction to reduce the number of predicted variables required for the prediction of a spectrum, while minimizing the effects of the noise in the training with the measured spectra. We found the best results by keeping only the first 20 principal components out of the 350 spectral components measured by the spectrometer.

We used multiple models to predict each of the output variables from the scaled features, and evaluated them using the mean error, calculated as the mean distance of each predicted value to the measured value. The training was performed to minimize the mean error on the training set. The hyperparameters of each model were modified to minimize the mean error on the validation set. Finally, the accuracy of each model was quoted as the mean error obtained on the test set. In the case of the spectral shape prediction, we define our accuracy by calculating the agreement between the vectors representing the measured, $V_{\textrm{m}}$, and the predicted, $V_{\textrm{p}}$, spectra using the similarity function defined as:
\begin{equation}
\textrm{Agreement} = \frac{2\mathopen|V_{\textrm{m}}\cdot V_{\textrm{p}}\mathclose|}{\mathopen|V_{\textrm{m}}\mathclose|^2+\mathopen|V_{\textrm{p}}\mathclose|^2} \textrm{ .}
\label{EqAgree}
\end{equation}

Polynomial models were fit to the data using simple regression. Due to the large number of features, it was not possible to use higher order models than quadratic, as the number of artificial features created by combining all of the input features up to the required degree scales as the number of k-multicombinations of n elements, where k is the polynomial order, and n the number of input features. In fact, the number of parameters to fit in the model can become comparable or larger than the size of the training data. In practice this limits the nonlinearities that can be represented, as the order is the only hyperparameter available to increase the complexity of polynomial models.

The optimal hyperparameters for the SVR models (C, epsilon, gamma) and the NN (number of hidden layers, number of cells per layer) were found in each case by applying a grid search. Each NN presented up to 3 hidden layers, with between 20 and 100 hidden cells per layer. A rectified linear activation function was used for the hidden cells. The NNs were trained until convergence using the Adagrad\cite{duchi2011adaptive} algorithm with a batch size of 1000 samples per training step. The final hyperparameters were chosen to minimize the error of the validation set, while not overfitting the training set, to make sure the model was kept as simple as possible. K-neighbors and Decision Tree Regressor models were also used, but in general achieved worse results for all the examples.

The same technique should be applicable to make predictions for every shot in new XFEL machines working at MHz, as the training, validation, and testing steps can still be performed at a low repetition rate (below 1 kHz). Nevertheless the accuracy of the predictions in this case may be different from the values shown here, as the hidden correlations exploited by the machine learning models may change in the new XFELs.

\section{Results and discussion}\label{Section3}

	We applied our technique to predict the photon energy, the spectral shape, and the pump-probe delay of x-ray pulses, which are the critical parameters in x-ray spectroscopy and time-resolved studies. These characteristics are predicted for both single pulse and double pulse configurations. For each of the predictions we optimized four different models: a linear model, a quadratic model, a support vector regressor, and a neural network. The results are summarized in Table \ref{TableSummary}.

\subsection{Single pulse photon energy prediction}\label{Section3a}
\begin{figure*}
	\centering
\includegraphics[trim=0cm 0cm 0cm 0cm,clip,width=0.9\linewidth]{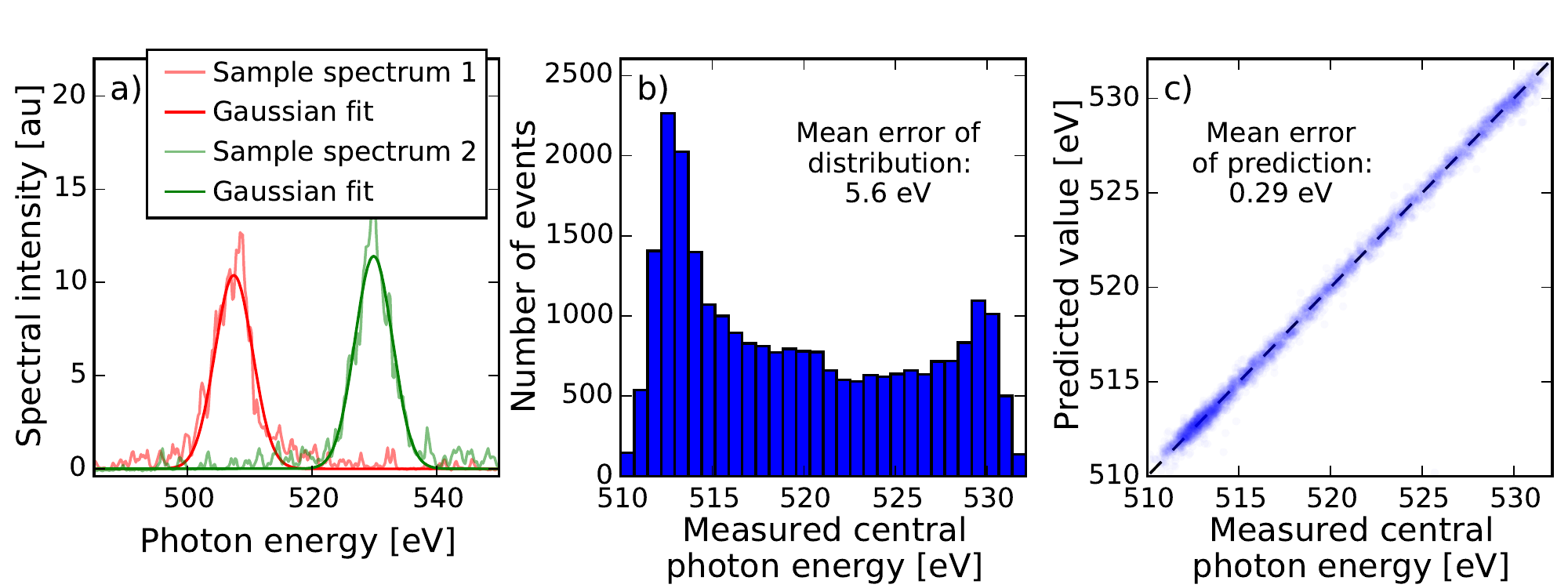}
	 \caption{\protect Photon energy prediction for a single pulse. (a) Two samples of single-shot spectra measured with the optical spectrometer at two different photon energies, and the corresponding Gaussian fits. (b) Distribution of the measured photon energies for the dataset. (c) Measured photon energies compared to the predicted photon energies for the test set using a linear model.}
	\label{FigSingle}
\end{figure*}

	For the calculation of the photon energy, we used the position of a Gaussian fit in our calibrated optical spectrometer as the variable to be predicted. Two examples of the experimental data with their corresponding Gaussian fits are shown in Figure \ref{FigSingle}a. The nominal electron beam energy was continuously scanned over a full range of 60 MeV (triangular periodic scan, 1 minute period, over 10 periods) to provide a wider range of photon energies for the predictions. The effect of this scan, combined with the inherent jitter, resulted in a distribution of photon energies spanning a FWHM of approximately 18 eV, corresponding to a mean error of 5.6 eV (Figure \ref{FigSingle}b).

The results show that all 4 models are able to predict the photon energy of the test set with a mean error of near 0.3 eV when compared to the actual measured values (Table \ref{TableSummary} and Figure \ref{FigSingle}c), reducing the error of the initial distribution by a factor of 20. While the error of the initial distribution was artificially enhanced by the electron energy scan, it is worth noting that the model is able to automatically pick up on its own and make use of the correlations between the variables due to the scan. Additionally, as the mean error due purely to the inherent jitter is near 1 eV, we still expect an improvement factor of near 3-4 in cases where the nominal electron energy is kept fixed.

These accurate predictions are not surprising due to the well known quadratic relationship between the electron beam energy and the photon energy given by the XFEL resonance condition. For small variations in energy (60 MeV change at 3500 MeV in our case), this leads to the linear approximation:
\begin{equation}
\frac{\Delta E_{\textrm{ph}}}{E_{\textrm{0,ph}}}=2\frac{\Delta E_{\textrm{e}}}{E_{\textrm{0,e}}}
\label{EqPE}
\end{equation}
where $E_{\textrm{0,ph}}$ and $E_{\textrm{0,e}}$ are the central photon and electron beam energies respectively, and $\Delta E = E - E_{\textrm{0}}$, where $E$ is the single-shot energy. In this way, the electron beam energy, measured non-invasively at the LCLS by an electron beam position monitor in the final dispersive section, can be used to sort data as a function of photon energy. On the other hand we observed that, if we train our models using the electron beam energy as the only feature, then the mean error achieved is still as high as 0.7 eV, and in fact it is necessary to include at least 20 features to achieve an error rounding to 0.30 eV. This suggests that, even in a simple case like this one, useful information about the photon energy is contained not just in the main feature, but it is also encoded in other variables.

Nevertheless, most of the correlations relevant to predict the photon energy seem to be essentially linear. As a consequence, the quadratic and the SVR models tend to overfit the data, showing a larger error for the test set than for the train set (Table \ref{TableSummary}). Similarly the best performance of the NN was obtained including only two hidden layers, with 10 and 5 hidden cells respectively, greatly limiting the degree of nonlinearities that the NN can model. While the degree of overfitting was not problematic for our purposes, regularization\cite{cheng1994neural} or dropout\cite{srivastava2014dropout} techniques could be applied to avoid it if necessary.

\subsection{Single pulse spectral shape prediction}\label{Section3b}
\begin{figure*}
	\centering
\includegraphics[trim=0cm -1.25cm 0cm 0cm,clip,width=0.35\linewidth]{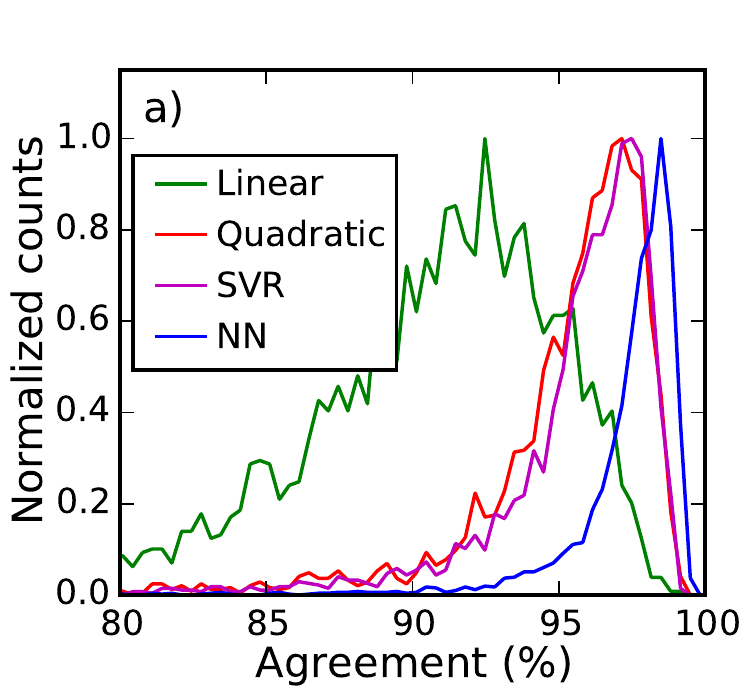}
\includegraphics[trim=0cm 0cm 0cm 0cm,clip,width=0.55\linewidth]{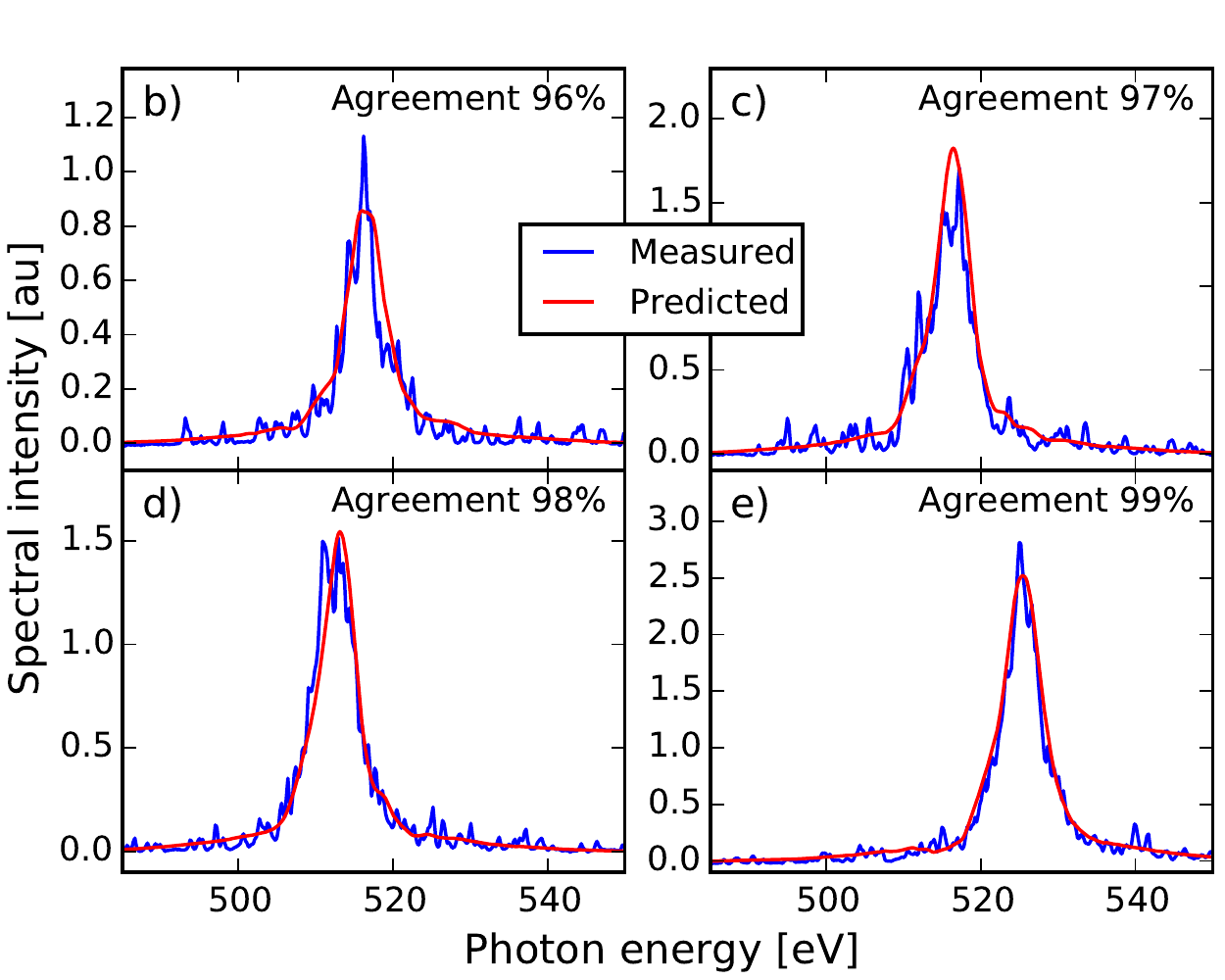}
	 \caption{\protect Spectral shape prediction for a single pulse. (a) Histogram of agreements between the predicted and the measured spectra for the test set using the 4 different models. (b-e) Examples of the measured and the predicted spectra using a neural network to illustrate the accuracy for different agreement values.}
	\label{FigShape}
\end{figure*}

In this case, instead of predicting the photon energy as a feature obtained from fitting the spectrum, we built models to directly predict the spectral shape by predicting multiple spectral components. The histogram of agreements between the measured and predicted spectra for the test set are shown in Figure \ref{FigShape}a. As this problem is much more non-linear than the previous case, the linear model only achieves a mean agreement of 88\%, while the other three models achieve mean agreements above 94\% (Table \ref{TableSummary}).

In particular the chosen NN, consisting of three hidden layers with 50, 50 and 20 hidden cells respectively, allows the network to find and model the non-linearities required for the prediction with a mean agreement of 97\%. In fact, 86\% of the shots in the test set show an agreement higher than 96\%. Figures \ref{FigShape}b-e show examples of predicted compared to measured spectra for increasing agreements from 96\% to 99\%. Even the example with the lowest agreement shows a good match, including more details of the spectral shape that can be achieved with a Gaussian or Lorentzian fit.

It is worth noting that due to the non-linearity of the problem, none of the models seem to overfit, making this a possible symptom of a \emph{high-bias} situation, meaning that given more training, more features, or more complex models, even better results could be achieved.

Apart from potentially solving the repetition rate problem, this technique could also be of interest in absorption experiments, where the spectrum after absorption through a sample has to be measured and compared to the reference spectrum. Normally the reference spectrum is measured before inserting the sample and averaged for many shots, or even averaged for shots sorted in different bins as a function of one or two of the features\cite{kimberg2016stimulated}. However, this approach cannot be used to bin with respect to more than two variables, as then the number of samples per bin would become too small. Instead a model could be trained to learn how to predict the reference spectrum based on training reference data obtained without an absorption sample, and then used to predict the incoming spectrum for each single-shot measurement with the sample, allowing the calculation of single-shot absorption. This approach could be successful as long as reference data are recorded sufficiently often to account for long term drift in the machine. 

\subsection{Double pulse time delay prediction}\label{Section3c}
\begin{figure*}
	\centering
\includegraphics[trim=0cm 0cm 0cm 0cm,clip,width=0.9\linewidth]{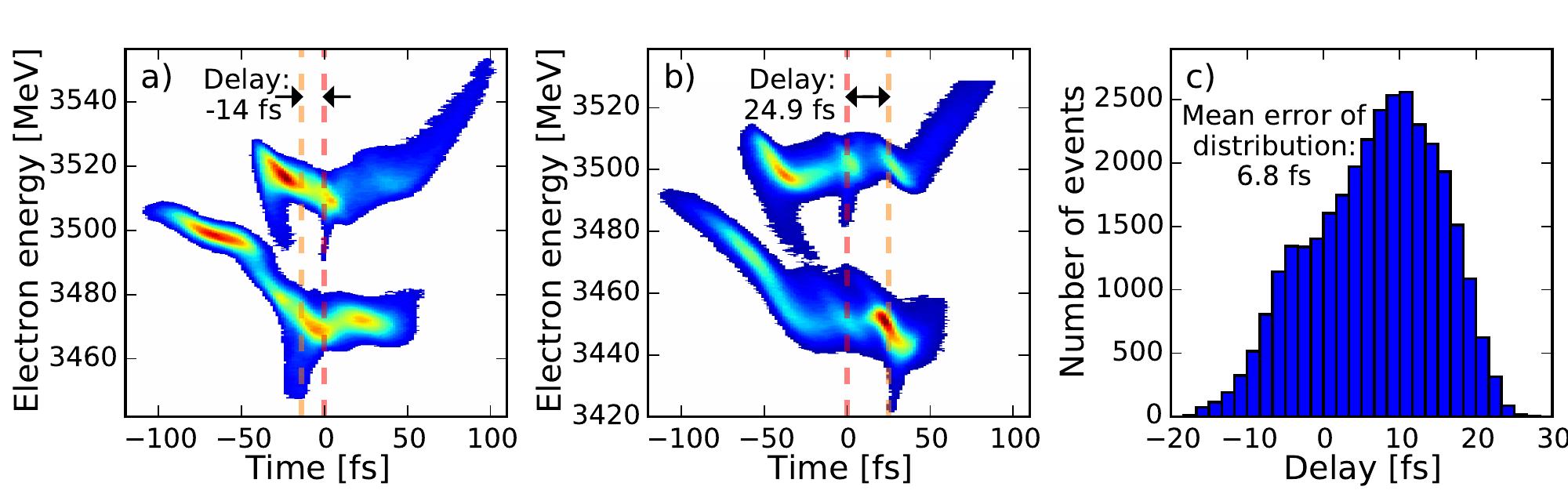}
	 \caption{\protect (a-b) Examples of the XTCAV traces used to extract the delay values by locating and measuring the distance between the lasing part of each bunch. (c) Distribution of all the delay values for the dataset.}
	\label{FigDelay1}
\end{figure*}
\begin{figure*}
	\centering
\includegraphics[trim=0cm 0cm 0cm 0cm,clip,width=0.55\linewidth]{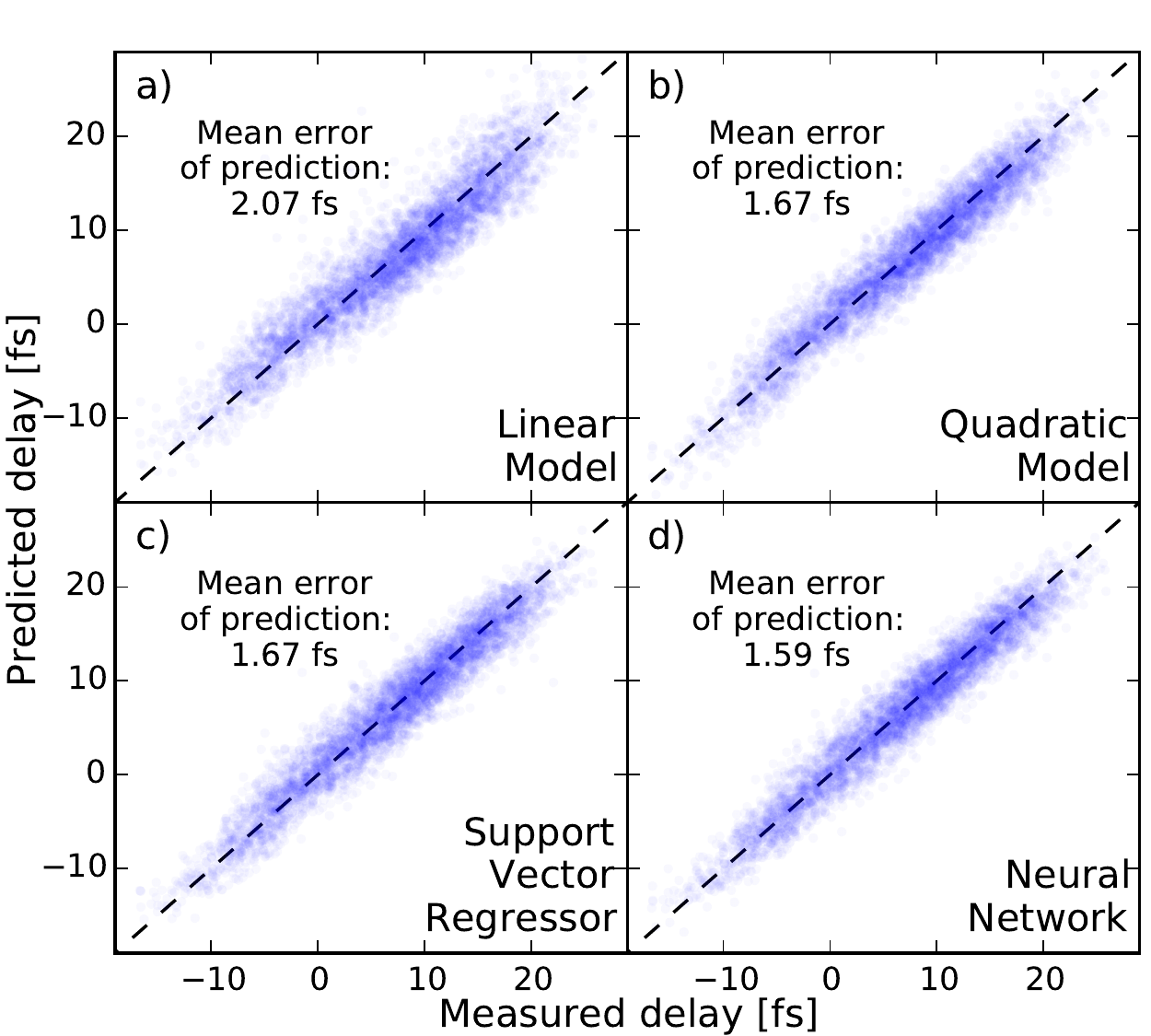}
	 \caption{\protect  Delay prediction errors for the test set using each of the four models.}
	\label{FigDelay2}
\end{figure*}
\begin{figure*}
	\centering
\includegraphics[trim=0cm 0.05cm 0cm 0.3cm,clip,width=0.55\linewidth]{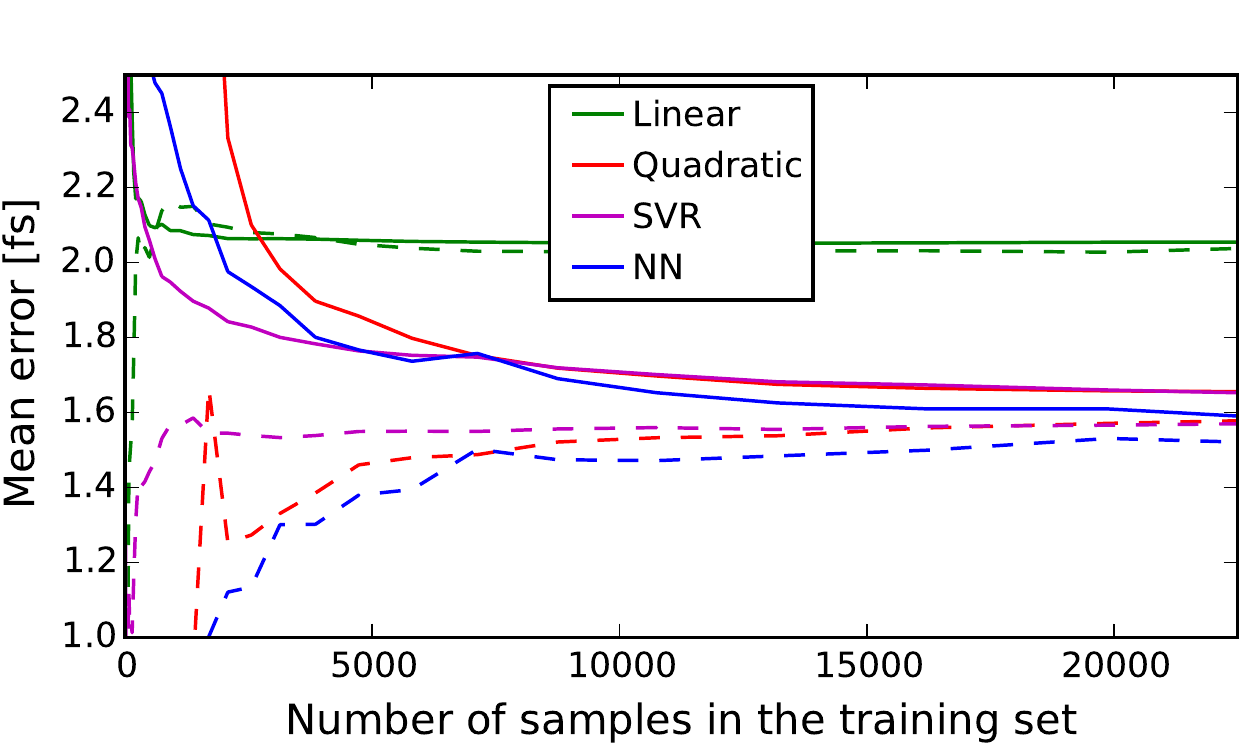}
	 \caption{\protect Delay prediction learning curve showing the mean error for the validation set (solid line), and the training set (dashed line) for each of the four models as function of the number of samples used for training.}
	\label{FigDelay3}
\end{figure*}

The delay values between the two x-ray pulses were extracted from electron time-energy distribution images recorded using the XTCAV diagnostic systems. Each image was processed by first separating the two bunches, and then locating the lasing part which appears as a temporally localized loss of electron beam energy and increase of energy spread when compared to non-lasing references\cite{ding2011femtosecond,behrens2014few,maxwell2014femtosecond}. Figures \ref{FigDelay1}a and \ref{FigDelay1}b show two XTCAV images, where the lasing slices have been highlighted with a red dashed line for the high energy bunch, and an orange dashed line for the low energy bunch. These two figures, obtained from the same dataset, already show two situations with opposite delay values. In fact, the distribution of the delays due to the jitter (Figure \ref{FigDelay1}c) spans  a FWHM of 25 fs, yielding a mean error of 6.8 fs.

After training all four models using the delay values from the training set, they were applied to the test set to predict the delay values. We found that all the models are able to predict the delay with a mean error near or below 2 fs (Table \ref{TableSummary}). Considering the mean error of the initial distribution was 7 fs, this already represents an improvement factor of at least 3.5 on the accuracy of the delay. 

\begin{figure*}
	\centering
\includegraphics[trim=0cm 0.05cm 0cm 0.5cm,clip,width=0.9\linewidth]{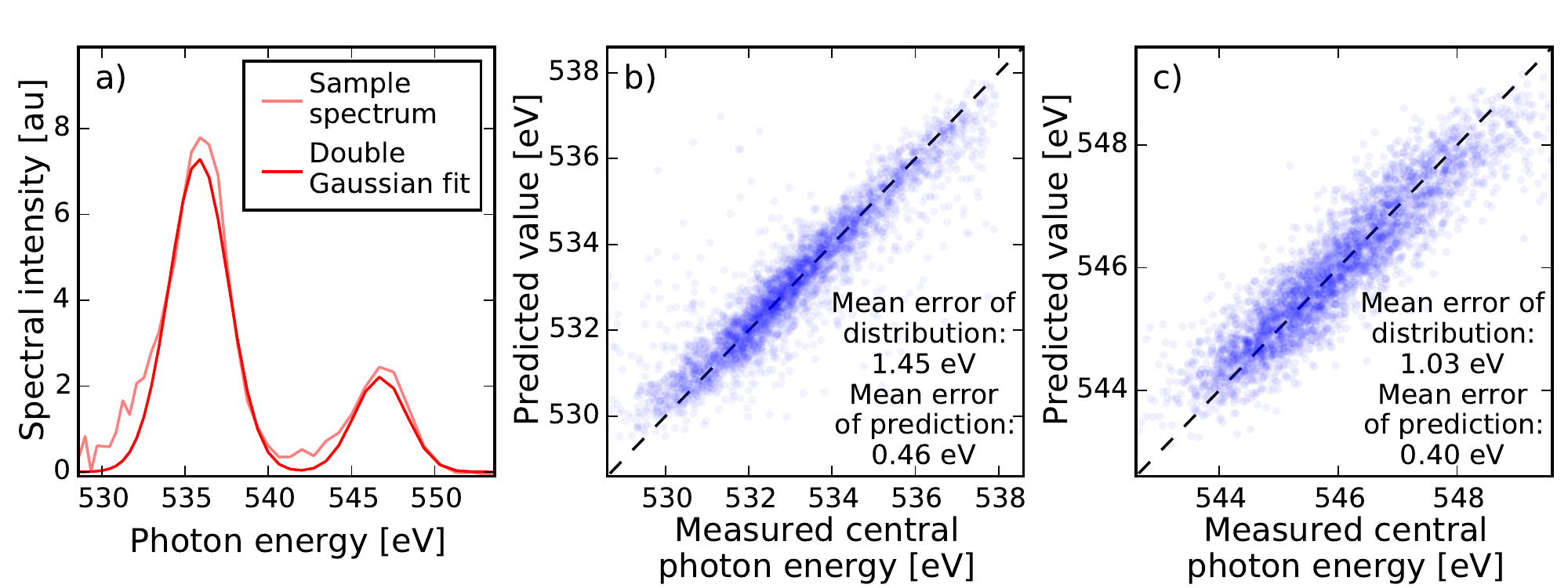}
	 \caption{\protect Description and results of the photon energy prediction for a double pulse mode. (a) A sample of a double pulse spectrum measured with the TOF spectrometer and the corresponding double Gaussian fit. (b,c) Measured photon energies of each of the pulses compared to the predicted photon energies for the test set using a NN.}
	\label{FigDouble}
\end{figure*}

As the physical processes that determine the final delay are complex, the non-linear models show better results, below 1.7 fs mean error. In particular, a NN with two hidden layers, with 50 and 10 hidden cells, predicts the delay with a mean error below 1.6 fs. From Figure \ref{FigDelay2}, we also observe that it is the neural network that presents the most symmetric deviation from the perfect correlation (dashed line), as opposed to the other models where there is some asymmetry in the residuals.

Most of the models (except the linear) seem to overfit, showing larger values for the error of the test set, than that of the training set (Table 1.). This could be a symptom of a \emph{high-variance} situation where the training could benefit from having more training data. In order to determine if this is the case we studied the accuracy of the predictions for the training set and the validation set as a function of the number of samples used for training (Figure \ref{FigDelay3}). This shows that, except for the linear model, all the other models have not fully converged to a value, so given more training data better results would be obtained, and maybe even more complex models could be fitted. Nevertheless, it is worth noting that all the non-linear models can predict the delay with a mean error smaller than 1.8 fs with only 5000 events which at a repetition rate of 200 kHz could be recorded in only 25 seconds.

While XTCAV is essential to measure some values of the delay, this result shows that it is possible to learn how to create models that calculate the delay from simpler parameters, which can be measured at a higher repetition rate. For an experiment aiming to measure few femtosecond dynamics, requiring XTCAV, this opens the possibility of actually recording data at the full repetition rate, not being limited by the XTCAV maximum repetition rate. This will be critical for the next generation of high-repetition-rate XFELs, but can also be retroactively applied to previous experiments at LCLS, where the XFEL and the data acquisition were working at 120 Hz, but XTCAV data was only recorded at 60 Hz.

\subsection{Double pulse photon energy prediction}\label{Section3d}

Following a similar approach as in section \ref{Section3a}, we used the electron TOF spectrometer in the double pulse mode to monitor the photon energy of each of the pulses (Figure \ref{FigDouble}a). We scanned the electron energy over a range of 20 MeV, yielding a distribution of photon energies with mean errors of 1.45 eV and 1.03 eV for each of the pulses. In this case, as in the single pulse case, we observe that all four methods show similar results (Table \ref{TableSummary}), with the NN (two hidden layers, with 20 and 5 hidden cells) yielding the smallest mean errors of 0.46 eV and 0.40 eV respectively (Figure \ref{FigDouble}b and \ref{FigDouble}c).

Nevertheless the absolute errors are still larger than the 0.3 eV mean error obtained for the single pulse. We believe the main reason for this is the lower signal-to-noise ratio of the TOF spectrometer, that we estimated to be 16 times worse than that for the optical spectrometer under the experimental conditions used. Furthermore the mean total x-ray energy was the same in both cases (\textasciitilde30 \textmu J), but in the double pulse mode each of the pulses carried only half the energy, providing less signal. As a consequence, the accuracy of the fits is reduced, giving less reliable values for the central photon energy.

Additionally, we attempted to perform the full spectral prediction in this case, but found that while the predicted spectrum matches well the position of the peaks, it does not predict the correct relative intensities between the two pulses. The first reason for that could again be related to the lower efficiency of the TOF spectrometer. Another possible reason is that, regardless of how much we learn about the trajectory of the bunches, the stochastic SASE emission cannot be easily predicted, as it depends on the microscopic structure of the bunch which is not yet possible to measure using existing diagnostics. In the single pulse mode this is not a problem, as the gas detector directly measures the total pulse energy for every single shot, however in a double pulse mode the gas detector cannot tell how much of the energy is in each of the pulses. All these considerations should be taken into account to better design future XFELs, by including simpler/faster diagnostics, placed strategically to have some correlations with the information we plan to predict, even if the correlations are not simple.

\section{Conclusion}\label{Section4}
We have shown that, for current existing data from LCLS, the fluctuations of the electron bunch trajectories measured with fast detectors encode important correlations with many of the required shot-to-shot x-ray properties, such as photon energy, spectral shape or time delay. Many of these  properties will not be available on a shot-to-shot basis at high-repetition-rate XFELs. In many cases, the critical properties for an experiment cannot be easily measured for all shots if the design of the experiment does not allow measuring downstream of the interaction region or the diagnostics require unfeasibly high data rate in high repetition operation.

We have presented a technique based on machine learning algorithms that allows many of these key shot-to-shot properties to be obtained, based solely on information from fast detectors recorded upstream of the interaction region, requiring only a small amount of training data that can be recorded for a subset of the shots, or at a lower repetition rate. This may even be used to automatically obtain shot-to-shot reference spectra for absorption measurements. In a more general application, the same method could also be applied to fill data gaps due to synchronization failures through the recording of a dataset, or even to perform online filtering of the events before the storage, which will be challenging at the MHz rates.

As test cases, we have successfully applied the proposed method for prediction of the photon energy (mean error \textless 0.3 eV at 540 eV), the spectral shape (mean agreement \textgreater 97\%), and the x-ray pulse delay (mean error \textless 1.6 fs) in a twin bunch mode, using four different machine learning models. We present the results from different models not to show a deep comparison of all 4 models, but rather to prove that when the necessary correlations exist, many machine learning models can exploit them, and even non-expert users should be able to apply the technique using the simpler and easier-to-train models.

We believe that putting together XFEL science with machine learning opens new opportunities, particularly for ultrafast time-resolved experiments, at new high-repetition-rate XFEL facilities, but also offers a new route to reanalyze data from past experiments, including experiments involving XTCAV or absorption experiments. Now that many aspects of the next generation of XFELs are being defined, this work provides evidence that the design of the new machines should not dismiss useful and difficult-to-replace complex diagnostics that cannot work at the full repetition rate, but should instead store as much full-repetition-rate single-shot information as possible and use the complex, low-repetition-rate diagnostics to complement them.

\section*{Acknowledgements}\label{Acknowledgements}
We acknowledge the support from Engineering and Physical Sciences Research Council (UK) (EPSRC) grant EP/I032517/1 and the European Research Council (ERC) ASTEX project 290467. A.S-G. is funded by the Science and Technology Facilities Council (STFC). H.F., K.M. and K.U. acknowledge support by the X-ray Free Electron Laser Utilization Research Project and the X-ray Free Electron Laser Priority Strategy Program of the Ministry of Education, Culture, Sports, Science and Technology of Japan.  R.F., V.Z. and J-E.R. would like to acknowledge the Swedish Research Council (VR). R.F., A.O.L and V.Z. would like to acknowledge financial support from the Knut and Alice Wallenberg Foundation (KAW), Sweden. V.Z. would like to acknowledge the Stockholm-Uppsala Center for Free Electron Laser Research, Sweden. M.I. acknowledges funding from the VW foundation within a Peter Paul Ewald-Fellowship. W.H. acknowledges financial support from a Marie Curie International Outgoing Fellowship. Use of the Linac Coherent Light Source (LCLS), SLAC National Accelerator Laboratory, is supported by the U.S. Department of Energy, Office of Science, Office of Basic Energy Sciences under Contract No. DE-AC02-76SF00515.
\vfill
\bibliography{references}

\clearpage
\section*{Appendix 1. Description of the variables used for the prediction}\label{Appendix1}
 
\subsection{Fast variables}\label{Appendix1a}
We list here all of the fast shot-to-shot variables used as features for prediction, currently measured at 120 Hz at LCLS: 
\begin{itemize}
\item ``ebeamCharge'' and ``ebeamDumpCharge'': Electron beam charge measured at the accelerators, and at the electron dump.
\item ``ebeamEnergyBC1'' and ``ebeamEnergyBC2'': Electron beam energy measured at each of the two bunch compressors.
\item ``ebeamPkCurrBC1'' and ``ebeamPkCurrBC2'': Electron beam peak current measured at each of the two bunch compressors.
\item ``ebeamLTUPosX'' and ``ebeamLTUPosY'': Horizontal and vertical electron beam positions at the Linac to Undulator (LTU) transport line.
\item ``ebeamLTUAngX'' and ``ebeamLTUAngY'': Horizontal and vertical electron beam angles at the Linac to Undulator (LTU) transport line.
\item ``ebeamLTU250'' and ``ebeamLTU450'': Electron beam position in two dispersive regions at the LTU transport line.
\item ``ebeamUndPosX'' and ``ebeamUndPosY'':  Horizontal and vertical electron beam positions at the undulator.
\item ``ebeamUndAngX'' and ``ebeamUndAngY'': Horizontal and vertical electron beam angles at the undulator.
\item ``f\_11\_ENRC'' and ``f\_12\_ENRC'': Redundant x-ray total energy measurements before attenuation from two gas detectors.
\item ``f\_21\_ENRC'' and ``f\_22\_ENRC'': Redundant x-ray total energy measurements after attenuation from two gas detectors.
\item ``f\_63\_ENRC'' and ``f\_64\_ENRC'': Redundant x-ray total energy measurements corrected to be accurate for small signals (\textless 0.5 mJ).
\end{itemize}

\vfill
\subsection{Slow EPICS variables}\label{Appendix1b}
We list here typical slow environmental properties recorded as EPICS variables measured at 2 Hz at LCLS: 
\begin{itemize}
\item Positions of translation stages involved in the control feedback loops.
\item Voltages of power supplies involved in the control feedback loops.
\item Strength of magnetic fields in the magnetic chicanes, and bending magnets.
\item Nominal values for the amplitude and phases of the radiofrequency fields.
\item Pressures from the vacuum systems.
\item Temperatures at different stages.
\item Calibration values inputted manually by operators.
\item Status of beam blockers.
\end{itemize}

\end{document}